%% file: main_withbib.tex
\documentclass[twocolumn,reprint,aps,prd,nofootinbib,floatfix,superscriptaddress]{revtex4-1}
 \usepackage[unicode]{hyperref} 
\usepackage[utf8]{inputenc}
\usepackage{amsmath}
\usepackage{amsfonts}
\usepackage{amssymb}
\usepackage[left=2cm,right=2cm,top=2cm,bottom=2cm]{geometry}
\usepackage{braket}
\usepackage{dsfont}
\usepackage{hyperref}
\usepackage{xcolor}
\usepackage[normalem]{ulem}

\usepackage{graphicx}
\usepackage{tikz}
\usetikzlibrary{arrows,decorations.pathmorphing,backgrounds,positioning,fit,petri,shapes}
\usepackage{lipsum}

\newcommand{\mnras}{Mon. Not. R. Astron. Soc.}
\newcommand{\apjl}{Astrophys. J. Lett.}
\newcommand{\jcap}{J. Cosmol. Astropart. Phys.}

\usepackage{multirow}
\begin{document}

\title{Learning Balanced Field Summaries of the Large-Scale Structure with the Neural Field Scattering Transform}

\begin{abstract}
We present a cosmology analysis of simulated weak lensing convergence maps using the Neural Field Scattering Transform (NFST) to constrain cosmological parameters. The NFST extends the Wavelet Scattering Transform (WST) by incorporating trainable neural field filters while preserving rotational and translational symmetries. This setup balances flexibility with robustness, ideal for learning in limited training data regimes. We apply the NFST to 500 simulations from the CosmoGrid suite, each providing a total of 1000 square degrees of noiseless weak lensing convergence maps. We use the resulting learned field compression to model the posterior over $\Omega_m$, $\sigma_8$, and $w$ in a $w$CDM cosmology. The NFST consistently outperforms the WST benchmark, achieving a $16\%$ increase in the average posterior probability density assigned to test data. Further, the NFST improves direct parameter prediction precision on $\sigma_8$ by $6\%$ and $w$ by $11\%$. We also introduce a new visualization technique to interpret the learned filters in physical space and show that the NFST adapts its feature extraction to capture task-specific information. These results establish the NFST as a promising tool for extracting maximal cosmological information from the non-Gaussian information in upcoming large-scale structure surveys, without requiring large simulated training datasets.

\end{abstract}

\author{Matthew Craigie}
\email{m.craigie@uq.edu.au}
\affiliation{School of Mathematics and Physics, The University of Queensland, QLD 4072, Australia}

\author{Yuan-Sen Ting} 
\affiliation{Department of Astronomy, The Ohio State University, Columbus, OH 43210, USA}
\affiliation{Center for Cosmology and AstroParticle Physics (CCAPP), The Ohio State University, Columbus, OH
43210, USA}

\author{Rossana Ruggeri}
\affiliation{School of Mathematics and Physics, The University of Queensland, QLD 4072, Australia}
\affiliation{School of Physics and Chemistry,  Queensland University of Technology, QLD 4000, Australia}

\author{Tamara M.\ Davis}
\affiliation{School of Mathematics and Physics, The University of Queensland, QLD 4072, Australia}

\maketitle


\section{Introduction}\label{sec:introduction}
\input{sections/1_introduction}

\section{Data \& Methodology}\label{sec:methods}
\input{sections/2_data}

\section{Results}\label{sec:results}
\input{sections/3_results}

\section{Discussion}\label{sec:discussion}
\input{sections/4_discussion}

\section{Conclusion}\label{sec:conclusions}
\input{sections/5_conclusion}


\section{Acknowledgements}
M.C. acknowledges the support of an Australian Government Research Training Program (RTP) Scholarship. M.C., R.R., T.M.D., acknowledge the support of an Australian Research Council Australian Laureate Fellowship (FL180100168) funded by the Australian Government. Y.S.T is supported by the National Science Foundation under Grant No. AST-2406729. R.R. acknowledges financial support from the Australian Research Council through DECRA Fellowship DE240100816. This research used resources of the National Energy Research Scientific Computing Center (NERSC), a U.S. Department of Energy Office of Science User Facility located at Lawrence Berkeley National Laboratory, operated under Contract No. DE-AC02-05CH11231.

\end{document}

%% file: sections/1_introduction.tex
The universe’s large-scale structure encodes rich information about cosmological parameters and their evolution. Effectively capturing this information is essential to maximize scientific returns from upcoming missions by tightening cosmological parameter constraints, in an effort to reconcile growing discrepancies between observations of the universe and the standard $\Lambda$CDM model of cosmology \citep{dainotti2023, camilleri2024, des2024sne, poulin2023, desi2025bao}. 

Extracting cosmological information from the large-scale structure requires effective compression of the various uninformative degrees of freedom in the field. For this compression, cosmology inference methods often use summary statistics: compact sets of statistics that capture selected information from the field. Traditional cosmology inference approaches develop analytic models that make predictions for how these statistics depend on cosmology, and compare these predictions to observations. The leading approach in modern large-scale structure survey analysis involves using two-point statistics, which are widely used across galaxy clustering \citep{masot2025}, weak lensing \citep{desy3shear, asgari2021, hscy1shear2020}, and joint clustering-lensing analyses \citep{desy3cross, heymans2021, hscy3}. 

However, these two-point methods cannot access the non-Gaussian portion of the large-scale structure, which is rich in cosmologically relevant information. Capturing this non-Gaussianity requires accessing higher-order information. This information is most directly accessed by $3$-point and $4$-point statistics \citep{gilmarin2015, gilmarin2017, slepian2017, philcox2021, ivanov2022, cagliari2025}, but modeling challenges and noise sensitivity limit their applicability to cosmology inference \citep{yankelevich2019, oddo2020, wang2024}.

One powerful alternative is to use deep learning to capture field information. Deep learning models are highly flexible, meaning they can adapt to capture cosmologically relevant information. The leading models in vision-based deep learning tasks, Convolutional Neural Networks (CNNs), show promise for constraining cosmology with 2D weak lensing datasets, with applications to the Kilo-Degree Survey \citep{fluri2019} and Dark Energy Survey \citep{jeffrey2024}. Likewise, CNNs have proven successful for constraining cosmology with the 3D galaxy distribution \citep{lemos2024}. 

However, CNNs generally require large datasets to generalize effectively due to their high flexibility. For cosmology simulations with sufficiently high precision to accurately model non-Gaussian field features, it can be a computational challenge to provide enough simulated volume to train effective CNN models. Instead, in this limited data regime, CNNs suffer from over-fitting and poor generalization. CNN models often address this issue through various forms of regularization. For example, \cite{jeffrey2024} use an ensemble of models and varied noise realizations in their training simulations to mitigate over-fitting. \cite{lemos2024} implement regularization in the form of dropout and an $\ell_2$ weight penalty, while \cite{roncoli2024} add a domain-adaptive penalty to the loss to improve generalization.

An alternative to these general regularization schemes is problem-specific, physically-informed regularization. Using assumptions about the data, known as inductive biases, models are constrained to learn more optimal solutions, helping prevent noise-dominated learned solutions that generalize poorly. The large-scale structure is well-suited for inductive biases due to firm assumptions from the cosmological principle: global rotational and translational symmetry.

The Wavelet Scattering Transform (WST) method incorporates these inductive biases in a CNN-like architecture, providing fixed convolution filters and processing with rotational and translational symmetry. The WST has emerged as a powerful alternative to two-point approaches, with wide application to cosmology inference \citep{valogiannis2022b, valogiannis2022a, valogiannis2024}. However, while the WST's architecture is highly robust, its use of fixed wavelet filters means it is susceptible to missing additional useful information, since it lacks the flexibility to adapt to a specific field. 

To introduce flexibility, \cite{gauthier2022} use a WST architecture with learnable wavelet parameters and see an improvement over the standard WST for image classification. Similarly, \cite{khemani2022} develop Learnable Wavelet Scattering Networks that optimize wavelet filter parameters using genetic algorithms, improving fault diagnosis tasks in engineering domains.

Extending this flexibility further, in \cite{craigie2025} we introduce the Neural Field Scattering Transform (NFST) which adds trainable convolution filters to the WST's architecture. Crucially, the NFST parametrizes the filters using a neural field, which promotes more smooth and stable convolution filters than pixel-wise kernels. The NFST's architectural inductive biases and filter smoothness enhance its data-efficiency, while its flexibility improves information capture. Together, these provide a balanced approach that improves over both the WST and a CNN model for detecting parity violation in large-scale structure-like data \citep{craigie2025}. This previous success of the NFST motivates its application to different fields and training objectives.

In this work, we apply the NFST to constrain cosmology using weak lensing as a tracer of the large-scale structure. In section \ref{sec:methods} we introduce the CosmoGrid weak lensing dataset, describe the summary statistics in greater detail and explain our methodology. In section \ref{sec:results} we show the results, and visualize the differences between a trained NFST and the WST. We place these results in context in \ref{sec:discussion}, before concluding in section \ref{sec:conclusions}.

%% file: sections/2_data.tex
\subsection{Data}
For large-scale structure data, we use CosmoGrid, a suite of simulations designed for cosmology inference applications that includes 2500 cosmological N-body simulations run across varied cosmology parameters, with the primary grid simulations varying over $\Omega_m$, $\sigma_8$, $w$, $H_0$, $n_s$ and $\Omega_b$. CosmoGrid provides noiseless full-sky weak lensing convergence maps across 4 redshift slices. The maps are provided in healpix format with \texttt{nside=1024}, and have been sampled to correspond to Dark Energy Survey (DES) tomographic bins. The simulations incorporate baryonic feedback effects through a post-processing baryonification model that modifies the dark matter-only density distribution to account for baryonic physics effects on the matter field structure.

To reflect realistic constraints on simulation resources, we intentionally conduct our analysis in a limited data regime. In practical cosmology applications, high-fidelity simulations capture the rich non-Gaussian structure more accurately. However, increasing the fidelity of simulations often requires reducing the total number of realizations due to the heavy computational cost of accurately simulating small scale structure. Therefore, we aim to identify summary statistics that remain effective when trained on a small number of simulations. For this analysis, we choose a subset of 500 cosmologies for training (400) and validation (100), selected at random. The remaining 2000 are used as a test set to benchmark each summary statistic's  performance. 

We project the Healpix maps into flat $10^\circ\times10^\circ$ patches using a Cartesian projection. This patch size means projection distortions are limited to sub-percent in area, and distortions manifest independent of cosmology and should therefore have minimal effect in a cosmology analysis. The resulting fields have a pixel size corresponding to $\sim4$ arcmin, approximately matching the resolution limit of the healpix maps. 

Although CosmoGrid provides whole-sky convergence maps for each cosmology, we choose to use 10 of the $100~\mathrm{deg^2}$ patches, to approximate the survey size of a $\sim1000~\mathrm{deg^2}$ photometric survey such as the Kilo-Degree Survey (KiDS, \cite{kuijken2019}). Increasing area (a possibility with future surveys such as Euclid \citep{euclid2024} and LSST \citep{rubin2019}) would have the effect of reducing the noise on each summary statistic. However, given that this data uses a perfect, systematic-free and noiseless reconstruction of the weak lensing field, the resulting constraint precision should only be interpreted for relative comparison purposes, not as an indication of the potential recovered precision in a observational survey.  

\subsection{Summary Statistics}
We test the ability of three summary statistics to extract cosmologically useful information from the weak lensing fields. Each summary statistic outputs a different set of coefficients that contain information about physical aspects of the field.

\subsubsection{The Wavelet Scattering Transform}
The standard Wavelet Scattering Transform (WST) extracts information from homogeneous fields by applying  convolution filtering across a set of complex wavelet filters with different scales and orientations. The method convolves the input field $\delta$ with filters $\psi_{jl}$ (indexed by scale $j$ and angle $l$) and computes scattering coefficients via spatial averaging:
\begin{align}
S^1_{jl} &= \langle{\left|\delta \star \psi_{jl} \right|}\rangle \\
S^2_{j_1l_1j_2l_2} &= \langle{\left|\left|\delta \star \psi_{j_1l_1} \right|\star \psi_{j_2l_2}\right|}\rangle
\end{align}
where $\star$ denotes convolution and $\langle \rangle$ denotes spatial averaging. 

For an alternative intuition, we can consider convolution filtering in the Fourier domain, where it is simply the product of the Fourier input field $\hat\delta$ and the Fourier filter $\hat\psi_{jl}$. In this view, the complex wavelets are instead smooth, directional band pass filters. The first pass of convolution isolates structure of the first filter's scale, while the second pass of convolution captures how the first structure clusters on the second filter's scale. 

By averaging in the spatial domain, the first order coefficients measure the amount of structure of scale $j$, while second order coefficients measure the clustering (of scale $j_2$) of structures (of scale $j_1$). Second-order coefficients with $j_1 < j_2$ are not computed, with the physical intuition that structures cannot cluster on scales smaller than themselves. For first order coefficients, we average over angle index $l$ to preserve rotational symmetry. For second order coefficients, we apply an angular averaging scheme that preserves rotational symmetry while retaining cosmologically relevant relative angle information by averaging over angular differences $\Delta_\ell = |l_1 - l_2|$ rather than individual angles. Through these physically-motivated symmetry assumptions, the WST produces a compact set of coefficients representing the clustering information in the field. 

For our tests, we use $J=3$ scales (maximum scale $\sim30$ arcmin) and $L=6$ angles, giving 4 $\Delta_\ell$ coefficients for second-order terms. Although going to larger scales would likely tighten cosmology constraints, this information starts to be dominated by linear structure, which can be completely characterized by the power spectrum, and capturing this information is therefore not the goal of this test. 

\subsubsection{The Neural Field Scattering Transform}
The Neural Field Scattering Transform (NFST) extends the WST by replacing fixed Morlet wavelet filters with trainable neural field filters \citep{craigie2025}. The filters are parameterized as $\hat{\psi}_{jl}(\vec{k}) = F_{jl}(\vec{k}, j)$, where $F_{jl}$ is a neural network that maps Fourier-space coordinates $\vec{k}$ to filter values. Filters learn directly in Fourier space since in practice the convolutions occur in this domain for computational efficiency. Different scale filters use the same neural network, but providing the scale index $j$ as an additional input enables some filter variation across scales. We apply Fourier space truncation such that the $j$th filter is cut at frequencies above $N/2^j$ (where $N$ is the field size), ensuring each filter processes different scales while providing computational speedup. The neural field parameterization allows trivial dilation and rotation operations to readily construct a full filter set, and additionally promotes filter smoothness in Fourier space leading to more stable information extraction. 

The NFST uses the same scattering architecture as the WST but with optimized filters that adapt to specific tasks. This combination of robust symmetry-exploiting architecture with flexible trainable components excels in scenarios with limited data but high complexity requirements. We provide full implementation details in \cite{craigie2025}. For this work, we use a neural field network with 64 neurons, initialized to Morlet wavelets. The NFST model requires more computational time to train its filters than a standard WST. However, efficient GPU implementations and Fourier space truncation speedups means the NFST converges in $<30$ minutes for this dataset on a single GPU. The result of this additional training time is a set of coefficients with equivalent scale and angle properties as the WST that are richer in task-specific information.

\subsubsection{Convolutional Neural Network Compression}
Convolutional Neural Networks (CNNs) are a popular class of deep learning architectures that extract spatial features from 2D images by convolving the input with a series of learnable filters. Unlike the WST and NFST, which incorporate symmetry constraints explicitly, CNNs learn their features purely from data, allowing them greater flexibility but at the cost of reduced robustness. 

CNNs are distinct from the WST and NFST summary statistics in that they do not have explicitly defined output coefficients. However, CNNs usually construct compact internal representations of the field, which act as pseudo-summary statistics. To provide an equivalent comparison, we construct the CNN to have an output dimension equivalent to the number of coefficients in the WST and NFST models, effectively producing a set of CNN coefficients.  

The CNN architecture begins with a three-stage convolutional block, consisting of a $3{\times}3$ convolution with 32 output channels, stride 1, and padding 1, followed by ReLU activation, $2{\times}2$ max pooling, and dropout regularization. The resulting feature maps are flattened and passed through a two-layer fully connected latent embedding, comprising a ReLU-activated linear layer followed by dropout and another linear projection to an output vector that matches the dimension of the WST and NFST outputs, of size 16. 

This CNN architecture represents a standard baseline that is appropriate for the data volume and complexity of our cosmological inference task. While more sophisticated CNN architectures exist, they would likely suffer from overfitting given our dataset size. By using a relatively simple CNN architecture and channel-based and feature-based dropout regularization, we make the CNN more robust to over-fitting, thereby providing a fairer comparison to the WST and NFST models. 

For each statistic, we average computed coefficients over the ten field patches per cosmology, resulting in one set of 16 coefficients per redshift bin, for a total of four sets of coefficients. We concatenate these and normalize across cosmologies to provide the final set of 64 coefficients per cosmology for each summary statistic.

Together, these three summary statistics provide a spectrum of balance between informative power and incorporated symmetries, providing increasing flexibility and decreasing robustness in the order they are listed. In these tests, we do not compare to an explicit power spectrum summary statistic, since several previous studies have already demonstrated that the WST's capabilities exceed the power spectrum for capturing information in the non-linear large scale structure (e.g. \cite{cheng2021weak, valogiannis2022a}).

\subsection{Cosmology Inference}

We benchmark the performance of the three summary statistic models for predicting cosmology parameters from the large-scale structure. We train each summary statistic model to predict $\Omega_m$, $\sigma_8$ and $w$ from the CosmoGrid weak lensing datasets, demonstrating their ability to constrain $w$CDM cosmology. The other cosmology and astrophysical parameters varied in CosmoGrid are unconstrained by the model, and are therefore automatically marginalized over.

The full posterior modeling process for each summary statistic comprises three stages:
\begin{enumerate}
    \item Summary Statistic Pre-Training: Use raw lensing fields to predict cosmology parameters with the NFST and CNN models. Then, freeze the models' trainable parameters such that they behave deterministic summary statistics rather than trainable models, and compute the corresponding set of coefficients for each.
    \item Direct Parameter Prediction: Use explicit coefficients to directly predict cosmology parameters.  
    \item Posterior Modeling: Use predicted cosmology parameters to construct a posterior probability density model. 
\end{enumerate}

\subsubsection{Summary Statistic Pre-Training}

The NFST and CNN models contain trainable parameters that are tuned throughout training, while the WST model is fixed and deterministic. To treat each summary statistic equally in the downstream analysis, we choose to pre-train the NFST and CNN summary statistics and freeze their trainable components. Each model then acts as a fixed function that produces an equally-sized set of coefficients, rather than an end-to-end trainable model.

Another advantage of this approach is the ability to use a different training setup for each phase. In the pre-training phase, we train each model for fewer epochs but at a higher learning rate to quickly learn useful feature representations. Then, in the downstream training phase, we train the final prediction model for more epochs at a lower learning rate to fine-tune the cosmological parameter predictions. We find that this two-stage approach consistently achieves higher performance with significantly lower compute time requirements compared to end-to-end training.

For pre-training, we connect each model's output coefficients to a basic Multi-Layer Perceptron (MLP) neural network with an output for each cosmology parameter. We then train the models to predict cosmology by minimizing the mean-squared error loss between predictions and target cosmology parameters. The MLP model consists of two linear layers with 512 nodes and ReLU activation, and a final linear layer feeding to an unconstrained, 3-dimensional output corresponding to the cosmology parameters. We pre-train for 300 and 500 training epochs (iterations over the full training dataset) for the NFST and CNN models respectively, with an Adam optimizer using a learning rate of $10^{-3}$. For each model, the chosen number of training epochs is sufficient to achieve convergence, as determined from the train and validation loss curves. For each, we use early stopping to ensure generalizable solutions. After pre-training, we freeze the trainable components in each model and apply the models to each weak lensing field, computing their corresponding set of 64 coefficients each. 

\subsubsection{Direct Parameter Prediction}
We next train a regression model to predict the respective cosmology parameters from each of the summary statistics' coefficient sets, again by minimizing the mean-squared error loss. We use the same MLP model and training setup as with pre-training, but instead train for 2000 epochs with a learning rate of $10^{-4}$. 

After training, we test performance by computing the Root Mean Squared Error (RMSE) between predicted cosmology and true cosmology in the test set, providing a measure of prediction accuracy in physical units. A lower RMSE indicates that the summary statistic can provide tighter constraints on that cosmology parameter. 

Deep learning models have inherent stochasticity which introduces variability to test performance. This variability is amplified in scenarios with limited training data, where over-fitting is more prominent. Since we are interested in assessing the summary statistics' performance, we repeat each test 10 times to confirm that any improvement is systematic and not a stochastic artifact. 


\subsubsection{Posterior Modeling}


While predicting cosmology parameters directly is a useful indicator of summary statistic constraining power, it fails to capture the uncertainty in predictions and any degeneracies between parameters. To capture these, we instead use Neural Posterior Estimation (NPE). NPE is an approach where a neural model learns to approximate the posterior distribution of parameters given observed data, without requiring explicit likelihood evaluation. NPE trains a neural density estimator to directly model $P(\theta|x)$ by learning from pairs of simulated parameters $\theta$ and corresponding observations $x$. It works by training a conditional density estimator to learn the mapping from data to posterior distributions. Once trained, we can sample from the NPE model to provide full posterior estimates for new observations, capturing both parameter uncertainties and degeneracies that point estimates cannot reveal \citep{cranmer2020}. As observations $x$ we use the predicted cosmology parameters from the previous direct parameter estimation. Using these highly-compressed parameter predictions removes the compression task from the density model, which we find results in more accurate posteriors than using the raw coefficients directly.

To fit the NPE model, we use the \texttt{sbi} python package \citep{tejerocantero2020} and opt to use a Masked Autoregressive Flow (MAF, \cite{papamakarios2018}) as a density model. For this setup, we perform a complete grid search over \texttt{hidden\_features} from [16, 32, 64, 128] and \texttt{num\_transforms} from [2, 4, 6, 8, 10, 12], finding that \texttt{hidden\_features=64} and \texttt{num\_transforms=8} produce the most accurate posterior. We emphasize that we are interested in the relative performance between the models under different input coefficients, so optimizing the posterior model further is secondary to ensuring consistency between tests. Within \texttt{sbi}, we use an uninformative flat prior over $[-5, 5]$ for each of the normalized parameters. CosmoGrid's finite cosmology parameter space sampling imposes an effective prior, but this is equivalent across all summary statistics, and all test cosmologies fall within this prior. 

We assess the effectiveness of each summary statistics' learned posterior by measuring their quality over the test set. As a metric, we compute the mean log probability across all data and parameter pairs in the test set. Higher values indicate the model assigns greater probability density to the true parameter values, with differences reflecting the relative quality of parameter estimation. This means that for higher values, the corresponding summary statistic provides a more accurate and trustworthy estimate of the cosmology parameters and their confidence. The 2000 data-parameter pairs form a sufficiently large test set to use mean log probability for assessing performance \citep{lueckmann2021}. Again, we report the results for 10 repeats to separate systematic improvement from training stochasticity.

%% file: sections/3_results.tex
\subsection{Cosmology Inference Models}

\begin{figure}
    \centering
    \includegraphics[width=1.0\linewidth]{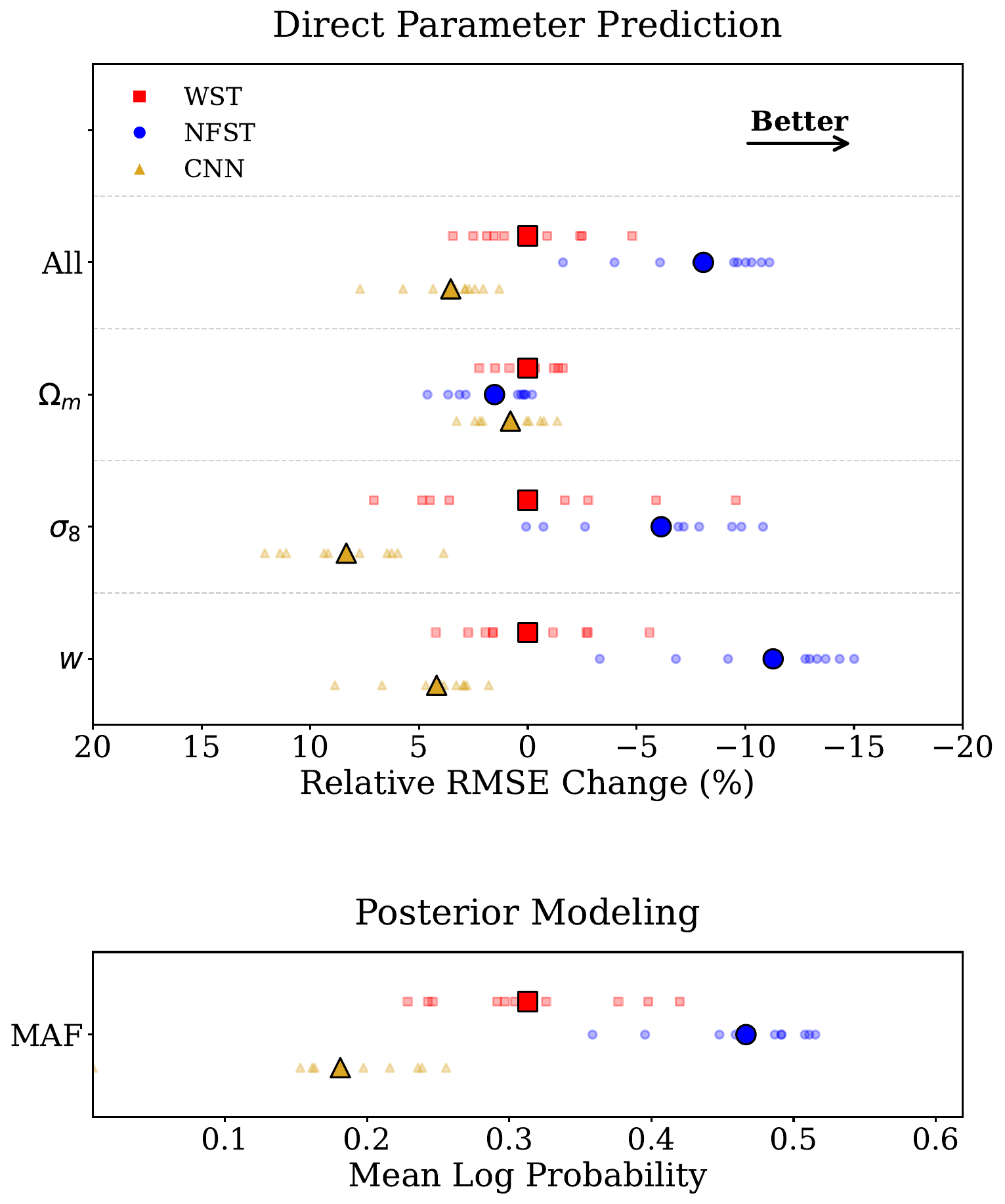}
    \caption{Comparison of the WST (red squares), NFST (blue circles) and CNN (golden triangles) summary statistics for constraining cosmology. Large shapes with black outlines indicate the average performance across 10 training runs, while smaller shapes indicate individual training runs. \textbf{Top:} Relative summary statistic performance for direct cosmology parameter prediction. We measure performance as the Root Mean-Squared Error (RMSE), relative to the average RMSE of the WST. \textbf{Bottom:} Mean log probability of a full posterior model that uses a Masked Autoregressive Flow (MAF) density estimator.}
    \label{fig:stat_comparison}
\end{figure}

To visualize the ability of the different summary statistics to capture cosmological information, in the top panel of Figure~\ref{fig:stat_comparison} we show the corresponding change in RMSE as percentages relative to the average RMSE of the WST model. To extend this to posterior estimation, in the bottom panel of Figure~\ref{fig:stat_comparison} we show the mean log probability for the posterior model built from each summary statistic. In both cases, we show the average performance across all ten repeats, as well as the individual runs to illustrate the model convergence variability. The model performance is often spread in a skewed or bimodal manner, but the average still provides a reasonable summary of the performance tendency of each model. 

In the direct parameter estimation test, the NFST consistently improves over the the WST. The NFST averages a 6\% decrease in prediction RMSE for $\sigma_8$, and a 11\% decrease in prediction RMSE for $w$. This improvement extends to posterior modeling, with consistently higher mean log probability over the unseen test data. Across the ten runs, the NFST averages a mean log probability of 0.47, compared to 0.31 for the WST. This corresponds to a $17\%$ increase in the average probability density assigned to each parameter-cosmology pair in the test dataset, which is a substantial improvement in posterior quality. These results show that the NFST recovers additional information compared to the WST from the small scales of the weak lensing convergence field. We see no improvement in $\Omega_m$, indicating that the WST already captures the majority of the field's information related to matter density. 

In contrast, the CNN summary statistic shows a consistent decrease in performance in all three cosmology parameters. Likewise, for posterior modeling, the CNN generally provides less accurate posterior constraints over the test set, equivalent to a 14\% decrease in average assigned probability density over the test set. This result indicates that the CNN struggles to capture information in this limited data regime, resulting in over-fitting and poor generalization to test data.

\subsection{Insights from NFST Filter Visualization}

A strength of the NFST coefficients is their direct physical meaning, with each coefficient representing the strength of features in the field with corresponding scale $j$ and angle $l$ (or relative angle $\Delta_l$ for second order coefficients) of the filters. In physical units, the three scales correspond to approximately 8, 16 and 32 arcmin respectively, which manifest as different physical scales for each redshift bin. 

\begin{figure}
    \centering
    \includegraphics[width=1.0\linewidth]{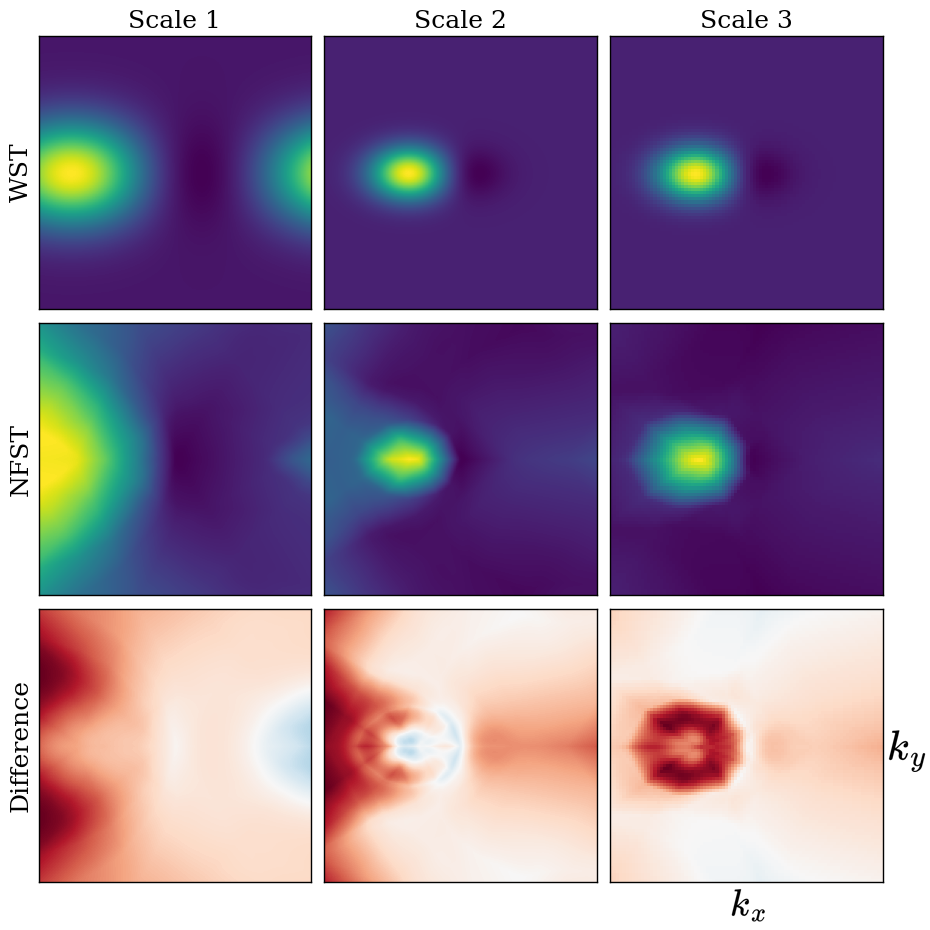}
    \caption{The first angular filter of the WST (top row) and trained NFST (middle row) shown in Fourier space (with coordinates $k_x$ and $k_y$), across the three scales used in the cosmology prediction. The middle of each image is the origin, with frequency increasing outwards. In the WST and NFST rows, brighter color indicates higher filter values. In the difference plot, red regions indicate an increase in the NFST's filters relative to the WST, while blue indicates a decrease. Scale 2 and Scale 3 appear visually similar despite representing different scales because the Scale 3 filter's high frequency modes are truncated for computational efficiency.}
    \label{fig:filter_compare}
\end{figure}

In Figure~\ref{fig:filter_compare} we visualize the differences between the learned filters and the standard WST's Morlet filters, to better understand the information that the NFST captures. 

The WST's Morlet filters' structure in Fourier space is a Gaussian bump centered on a nonzero frequency, with a smaller negative Gaussian bump subtracted at zero frequency to ensure the filter integrates to unity. The angle of this bump determines the associated angle in the filter. As a result, the Morlet filter acts as a smooth, directional band-pass filter. In each successive scale, the filter's highest frequency Fourier modes are removed, so successive filters capture larger scales. The first Morlet filter pushes up against the high-frequency edge, allowing it to capture the most small-scale information for each scale. To accommodate this, the Gaussian bump wraps to the other side to keep the wavelet smooth in configuration space.

The NFST's first filter, representing the smallest scales, exhibits a clear widening of the frequency pass band. This broader structure enables the downstream scattering transform operations to average over a greater number of high-frequency modes, likely improving the signal-to-noise ratio of the resulting coefficients. However, this comes at the cost of reduced precision on directional information. The preference for wide filters suggests that maximizing cosmological constraining power benefits more from averaging across broader frequency ranges than from maintaining precise angular resolution. The second and third filters both show a similar widening effect, though less pronounced. The third filter also displays a reshaped band-pass bump, shifting towards a sharper cutoff along both scale and angles.


\subsection{Insights from NFST Coefficient Visualization}
Although studying the NFST filters can provide insights into the first order coefficients and general areas of improvement, second order coefficients are more complex to understand. To help provide insights into the higher-order information that these coefficients access, we develop a method to directly visualize the information that the summary statistic captures. We then apply this to the NFST to visualize where the improvement over the WST originates from in physical, field-level space. 

\subsubsection{Summary Statistic-Based Field Maximization}
This new method for visualizing summary statistics relies on maximizing their value in a field. We expand upon work from \cite{cheng2021guide}, which visualizes the information that an arbitrary summary statistic captures from a target field. It works by minimizing the difference between the statistics evaluated over the target field and over a learned field. The minimization uses gradient descent, and can therefore leverage efficient deep learning framework computations and GPU acceleration. Building on this framework, we develop a new method that instead maximizes an individual summary coefficient over a learned field, while suppressing all other summary coefficients. This provides a clear visual representation of the information probed by that specific summary statistic.

Let $F(\theta)$ be the output field generated by a neural model with parameters $\theta$, and let $S(\cdot)$ be a vector-valued summary statistic function producing $k$ normalized components. We define normalization relative to a fixed baseline summary $S_{\text{base}}$, computed from a reference distribution, such that:
\begin{equation}
\hat{S}_i(\theta) = \frac{S(F(\theta))_i}{S_{i, \text{base}} + \varepsilon}
\end{equation}
where $\varepsilon$ is a small constant ensuring numerical stability. For a neutral baseline, we use uncorrelated standard normal noise. This normalization helps keep targets within the same order of magnitude, whereas each component of the summary statistic could be at any value. 

Let $I_C$ be the indices corresponding to the coefficients we wish to constrain towards 1 (thereby suppressing their characteristics in the field to the noise baseline), and $I_B$ for those we wish to boost beyond 1 (thereby enhancing their characteristics). We define the constraint potential $P_C(\theta)$ and the boost potential $P_B(\theta)$ as
\begin{equation}
P_C(\theta) = \left\langle \left| \hat{S}_i(\theta) - 1 \right| \right\rangle_{i \in I_C}
\end{equation}
and
\begin{equation}
P_B(\theta) = \left\langle \operatorname{sgn}\left(\hat{S}_j(\theta) - 1\right) \cdot \left| \hat{S}_j(\theta) - 1 \right| \right\rangle_{j \in I_B}
\end{equation}
where $\langle\ \cdot\ \rangle$ indicates an average over the respective set of indices. The use of the signed difference in $P_B$ prevents pushing towards highly negative values throughout optimization.

The total loss is defined as a weighted difference:
\begin{equation}
\mathcal{L}(\theta) = \lambda_C \cdot P_C(\theta) - P_B(\theta)
\end{equation}
where $\lambda_C \geq 0$ is a scalar parameter controlling the relative weight on the constraint potential. This loss penalizes components in $I_C$ for deviating from the neutral reference value of 1, and rewards components in $I_B$ for increasing above 1 and penalizes them for falling below it. Setting $\lambda_C=0$ promotes unconstrained maximization of $I_B$, which often results in non-convergent loss, but can still lead to useful visualizations after a moderate training period. 

Additionally, supplying a specific reference field and setting $I_C$ such that all indices are constrained is approximately equivalent to the method described in \cite{cheng2021guide}, providing a visualization of how a given summary statistic captures the information in a specific field. 

We also provide a modification to the trainable field. In \cite{cheng2021guide}, $F(\theta)$ is directly parameterized at the pixel level, such that each pixel is a trainable parameter. While using this approach in our new maximization framework proves effective for some summary statistics, for others it can result in convergence stuck at local minima, providing an incomplete representation of summary statistic properties.

To improve the optimization, we propose to parameterize $F(\theta)$ more flexibly with a dual-domain approach that operates in both real and Fourier space:

\begin{align}
F(\theta) = w_x \cdot [\mathcal{C}_x(z_x) + f_x] + w_k \cdot \mathcal{F}^{-1}[\mathcal{C}_k(z_k) + f_k]
\end{align}

where $\mathcal{C}_x$ and $\mathcal{C}_k$ are CNN models (we use U-Net architectures) that operate on learned latent spaces $z_x$ and $z_k$ respectively, $f_x$ and $f_k$ are additional learned fields in real and Fourier space, $w_x$ and $w_k$ are learnable weights controlling the relative contributions, and $\mathcal{F}^{-1}$ denotes the inverse Fourier transform. 

This dual parameterization provides complementary advantages: the real-space component $\mathcal{C}_x(z_x) + f_x$ excels at capturing localized features and sharp structures, while the Fourier-space component $\mathcal{F}^{-1}[\mathcal{C}_k(z_k) + f_k]$ naturally captures global patterns, periodic structures, and long-range correlations. The learnable weights $w_x$ and $w_k$ allow the optimization to automatically balance between these complementary representations based on the specific requirements of the target summary statistic.

This new parameterization is far more expressive than direct pixel-level optimization and has greatly improved convergence properties. The CNN formulation provides significant flexibility in both domains and can be made arbitrarily complex to provide better field statistic visualizations, at the cost of greater computation time.

Our method shares conceptual similarities with activation maximization techniques (e.g. \cite{erhan2009, zeiler2013, yosinski2015}), as both optimize a trainable input to enhance specific target outputs. While activation maximization seeks to amplify activations of particular neurons or layers exclusively within a neural network, our method instead generalizes beyond neural networks to visualize arbitrary summary statistics. 

\subsubsection{NFST Coefficient Visualization}

Using this new framework, we compare the NFST coefficient with the most influence on cosmology to its WST counterpart. To identify the most influential coefficients, we use the Shapley Additive Explanation (SHAP) framework, which explains the importance of a model's inputs (summary statistic coefficients) to its corresponding outputs (predicted cosmology parameter) \citep{lundberg2017}. 

The coefficient with the highest overall SHAP importance is the second order coefficient that corresponds to $j_1=1$, $j_2=2$ and $\Delta_\ell=0$, which physically represents how much $\sim4$ Mpc sub-structures tend to cluster on scales of $\sim16$ Mpc, where the sub-structures and their clustering align in the same direction. This coefficient has the highest SHAP value for both $\sigma_8$ and $w$. Interestingly, this coefficient represents the coupling between the two smallest scales probed by the NFST, indicating that the most relevant scales to cosmology are those that are deepest into the non-linear regime. 

\begin{figure*}
    \centering
    \includegraphics[width=1.0\linewidth]{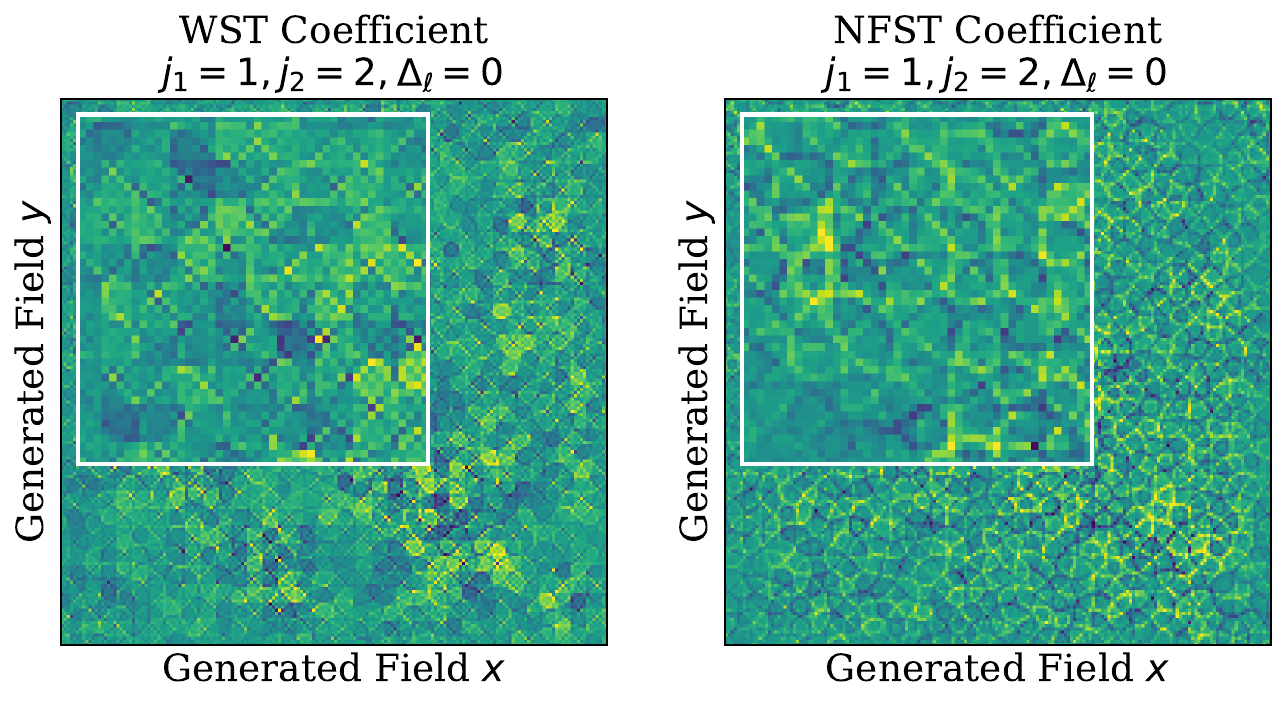}
    \caption{Generated fields representing the information probed by a selected second order scattering coefficient for the WST (left) and NFST (right) models. The white insets show a zoomed-in portion of the field. The selected coefficient is the most cosmologically-informative coefficient for the NFST. The generated fields show that the NFST extracts more filament-like information to better constrain $\sigma_8$ and $w$. We produce each field using the summary statistic visualization technique introduced in this work.}
    \label{fig:field_compare_first}
\end{figure*}

To visualize the information held by this coefficient, we show in Figure~\ref{fig:field_compare_first} the coefficient maximized for both the NFST and WST using the gradient descent technique discussed above. The visualization clearly shows that the NFST's learned filter capture a less rigid structure, reflecting the NFST's reduced angular precision in both the first and second order coefficients. This visualization presents a configuration space representation of the finding that the NFST aggregates information over more angular modes of information than its WST counterpart, complementing the Fourier space representation identified from the NFST's filters. The less rigid structures captured by the NFST more accurately reflect the chaotic and disorganized nature of the large-scale structure, and bears a stronger resemblance to filamentary structure. 

Beyond this specific application to the NFST, we demonstrate that this new method generally provides a useful and physically interpretable visualization of field summary statistics. Even for highly complex statistics with computations in Fourier space like the NFST, this approach can show which information the statistic is probing in physical space. 




%% file: sections/4_discussion.tex
The NFST's performance reveals a fundamental trade-off in cosmological inference: while maximum flexibility (as in CNNs) allows capturing arbitrary patterns, it requires prohibitively large training datasets in cosmology where simulations are computationally expensive. Conversely, fixed summary statistics like the WST impose strong prior assumptions that may discard cosmologically relevant information. The NFST demonstrates that incorporating physics-motivated constraints (translational and rotational symmetries) while allowing smooth task-specific filter adaptation provides more effective regularization than general deep learning techniques in data-limited regimes.

The learned filters' broader frequency pass bands suggest that cosmological information at small and intermediate scales is better extracted through statistical averaging across wide angular bands rather than precise spatial localization. This is a strong indication that fine angular resolution should not be prioritized in wavelet-based cosmology inference, instead referencing the decrease in statistical variance that comes through averaging over more modes.

Systematics such as survey masking and distance uncertainties can have a strong influence on cosmology inference accuracy with deep learning models \citep{desanti2025}. Furthermore, when moving towards observational datasets, the addition of noise can dampen non-Gaussian information, potentially reducing the information gain from the NFST. This contamination may exacerbate the need for wider filters to constrain statistical noise. The full effects of systematics and noise on the NFST's training procedure and inference accuracy warrants further investigation. 

While this work focuses on understanding how the NFST's captured information contrasts with the WST, future work can also explore a full interpretability analysis to identify the specific redshift bins and structural information most relevant to predicting each cosmology parameter. Such an analysis can provide unique insights into the connections between the large-scale structure and cosmology parameters. 

Future work can also explore the further application of the NFST to 3D large-scale structure, such as the full 3D galaxy distribution. Applications of CNNs to 3D LSS data have shown some success, but they rely on heavy regularization to avoid overfitting \citep{lemos2024}. Following its demonstrated robustness to over-fitting in 2D scenarios, the NFST may also extend well to 3D. With its physically-motivated inductive biases and smooth neural field filter representation, the NFST may provide the necessary advantage to flexibly capture 3D information from the field, without requiring excessive simulated training datasets. Such an extension to the NFST would require adaptation to match the line-of-sight asymmetry in observational large-scale structure, but could prove highly beneficial to improving cosmology constraints or better understanding the non-Gaussian large-scale structure.




%% file: sections/5_conclusion.tex
We demonstrate the use of the Neural Field Scattering Transform (NFST) as a summary statistic for large-scale structure. The NFST's balance between flexibility and robustness lets it achieve more accurate and precise cosmology constraints for cosmological inference compared to previous methods. Through trainable filters, the NFST provides an improvement in informative capacity relative to a standard Wavelet Scattering Transform (WST) model. For direct parameter prediction, the NFST provides a $6\%$ improvement in prediction RMSE for $\sigma_8$ and an $11\%$ improvement for $w$. For posterior modeling, the NFST provides a 17\% improvement in posterior quality. 

To maintain robust generalization even with limited training data, the NFST uses physically-motivated inductive biases, setting it apart from less physical Convolutional Neural Network (CNN) models. For direct parameter prediction, the NFST provides a $15\%$ and $17\%$ improvement in prediction RMSE for $\sigma_8$ and $w$ respectively, and  a $33$\% improvement over the CNN for posterior modeling. 


We also show the direct interpretability of the NFST by comparing it to the WST at the filter-level. We see that for the lensing field, Fourier-space filters with a broader angular spread are more effective for constraining cosmology. Physically, this suggests that for cosmology inference, directional information is less important than aggregation for reducing noise. 

To enable visualization at the coefficient level, we introduce a new summary statistic visualization technique. We use this to demonstrate how the NFST differs from the WST, and confirm in physical space that the NFST captures broader angular information, which more closely resembles filamentary structure. 

This new technique can also be applied beyond the NFST, to visualize other summary statistics. By enabling interpretability with this new method, we can verify the trustworthiness of deep learning-based statistics and provide physical insights into the data. 

More broadly, the NFST's ability to adapt to the specific lensing field suggests a domain-adaptability that could successfully capture information from other homogeneous fields beyond the cosmological large-scale structure. This work also marks the NFST's first application to cosmology inference, highlighting its task-adaptability beyond earlier demonstrations for parity violation detection. 

In future, we will apply the NFST to more realistic simulated data, testing its robustness to simulation artifacts and observational systematics, and ultimately apply it to weak lensing and clustering datasets from next-generation surveys. We also plan to extend the NFST to 3D, to efficiently and robustly capture richer 3D clustering information.

Overall, these findings demonstrate that flexible and robust models such as the NFST, combined with deep learning interpretability techniques, can improve the cosmological analysis process and deepen our understanding of large-scale structure.
